\documentclass[a4paper]{article}
\usepackage{INTERSPEECH2021}
\usepackage{amsmath,graphicx}
\usepackage{multirow}
\usepackage{makecell}
\usepackage{graphicx,color}
\usepackage{comment}
\usepackage{bbm}
\usepackage{color,colortbl}
\usepackage{xcolor}
\usepackage{adjustbox}
\usepackage{enumitem}
\usepackage{subfigure}
\usepackage[linesnumbered, ruled, vlined, algo2e]{algorithm2e}
\usepackage{array}
\usepackage{caption}
\usepackage{soul}
\usepackage{makecell}
\usepackage{booktabs}
\usepackage[colorlinks=true,allcolors=blue]{hyperref}
\usepackage[numbers,sort&compress]{natbib}

\definecolor{LightBlue}{RGB}{233,243,250}
\definecolor{LightRed}{RGB}{250, 214, 208}
\definecolor{Gray}{gray}{0.9}

\newcommand{\eat}[1]{}

\newcommand{\EMBR}{MWER\xspace}
\newcolumntype{g}{>{\columncolor{Gray}}c}
\usepackage{xspace}

\makeatletter
\DeclareRobustCommand\onedot{\futurelet\@let@token\@onedot}
\def\@onedot{\ifx\@let@token.\else.\null\fi\xspace}
\def\eg{\emph{e.g}\onedot} 
\def\ie{\emph{i.e}\onedot} 
 
 \def\vs{\emph{vs}\onedot}

\makeatother
\usepackage{amssymb}
\usepackage{amsmath,amsfonts}
\usepackage{amsopn,bm,mathtools}

\usepackage{nicefrac}       
\newcommand{\vct}[1]{\boldsymbol{#1}} 

\newcommand{\ProbOpr}[1]{\mathbb{#1}}

\newcommand{\expect}[2]{%
\ifthenelse{\equal{#2}{}}{\ProbOpr{E}_{#1}}
{\ifthenelse{\equal{#1}{}}{\ProbOpr{E}\left[#2\right]}{\ProbOpr{E}_{#1}\left[#2\right]}}} 
\newcommand{\var}[2]{%
\ifthenelse{\equal{#2}{}}{\ProbOpr{VAR}_{#1}}
{\ifthenelse{\equal{#1}{}}{\ProbOpr{VAR}\left[#2\right]}{\ProbOpr{VAR}_{#1}\left[#2\right]}}} 

\newcommand{\vx}{{\vct{x}}}
\newcommand{\vy}{\vct{y}}

\newcommand{\sL}{\mathcal{L}}

\title{Input Length Matters: Improving RNN-T and MWER Training for \\ Long-form Telephony Speech Recognition}

\name{
\begin{tabular}{c}
      Zhiyun Lu, Yanwei Pan, Thibault Doutre, Parisa Haghani, Liangliang Cao, \\
      Rohit Prabhavalkar, Chao Zhang, Trevor Strohman
      \end{tabular}
}
\address{Google Inc., USA}
\email{
\{zhiyunlu,llcao\}@google.com}

\begin{document}

\ninept
\maketitle
%




\begin{abstract}
End-to-end models have achieved state-of-the-art results on several automatic speech recognition tasks. However, they perform poorly when evaluated on long-form data, \eg, minutes long conversational telephony audio. One reason the model fails on long-form speech is that it has only seen short utterances during training. In this paper we study the effect of training utterance length on the word error rate (WER) for RNN-transducer (RNN-T) model. We compare two widely used training objectives, log loss (or RNN-T loss) and minimum word error rate (MWER) loss. We conduct experiments on telephony datasets in four languages. Our experiments show that for both losses, the WER on long-form speech reduces substantially as the training utterance length increases.  The average relative WER gain is 15.7\% for log loss and 8.8\% for MWER loss. When training on short utterances, MWER loss leads to a lower WER than the log loss. Such difference between the two losses diminishes when the input length increases. 

\end{abstract}

\noindent\textbf{Index Terms}:
end-to-end, long-form, telephony speech recognition, RNN-T, MWER
%

\section{Introduction}\label{sec:intro}

Automatic speech recognition (ASR) on telephony speech~\cite{tuske2021limit,xiong2018microsoft,han2017deep} is an important problem with real-world applications: \eg, medical conversations and call centers. 
End-to-end (E2E) models~\cite{zhang2020pushing} have achieved state-of-the-art performances on many benchmarks~\cite{gulati2020conformer}.
However, when we apply E2E models on the telephony conversations, its long-form nature and noisy acoustic conditions present a big challenge for the models' generalization capability and robustness~\cite{narayanan2019recognizing,chiu2021rnn}. In this paper, we investigate methods to improve E2E models performance on long-form telephony speech recognition. We focus on the RNN transducer (RNN-T)  model~\cite{graves2012sequence,Graves2013,JinyuLi2019,Zeyer2020,Saon2021}. 


One difficulty of the telephony speech recognition is that the audio is long-form. The recording length ranges from 30 seconds to more than 10 minutes.~\cite{chiu2021rnn,chiu2019comparison} proposed to segment the audio into short utterances before running inference. However, the segmented utterance would lose useful context information~\cite{hori21b_interspeech}, \eg, the speaker or topic, and the imperfect segmentation can introduce additional segmentation errors into the system. To correct the errors introduced by the segmentation, special handling like overlapping inference~\cite{chiu2019comparison} is needed in decoding. 
This is not ideal as it makes the system more complex and increases \eat{computational} latency. In this work, we apply E2E models \emph{without any segmentation} on the long-form data at test time. The model is able to leverage the context information, and the system is simple without any change in the infrastructure. 

\begin{figure}[tb]
    \centering
    \begin{tabular}{cc}
    \hspace{-0.3in}
    \includegraphics[width=0.26\textwidth]{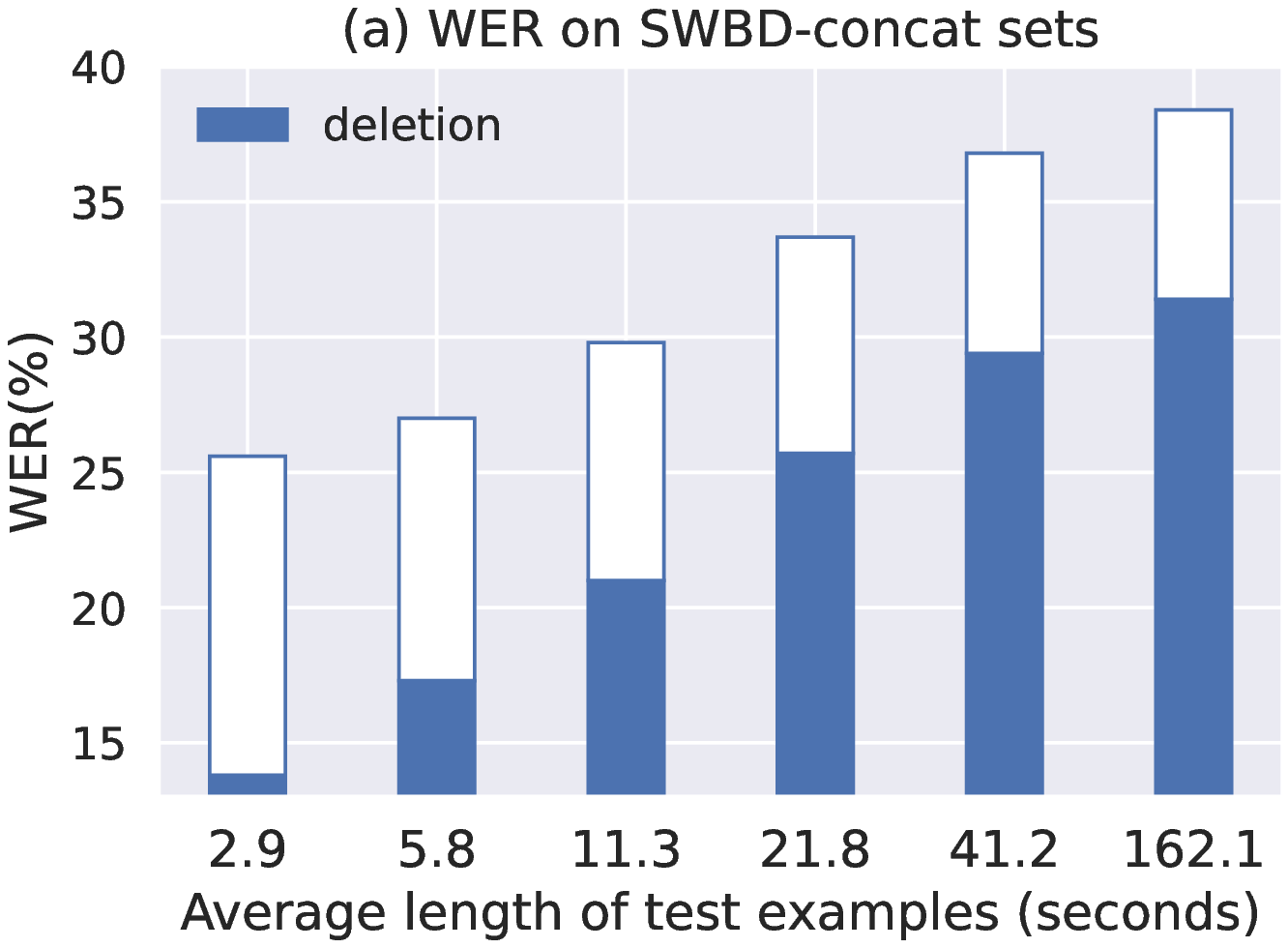}
    & \hspace{-0.2in} \includegraphics[width=0.248\textwidth]{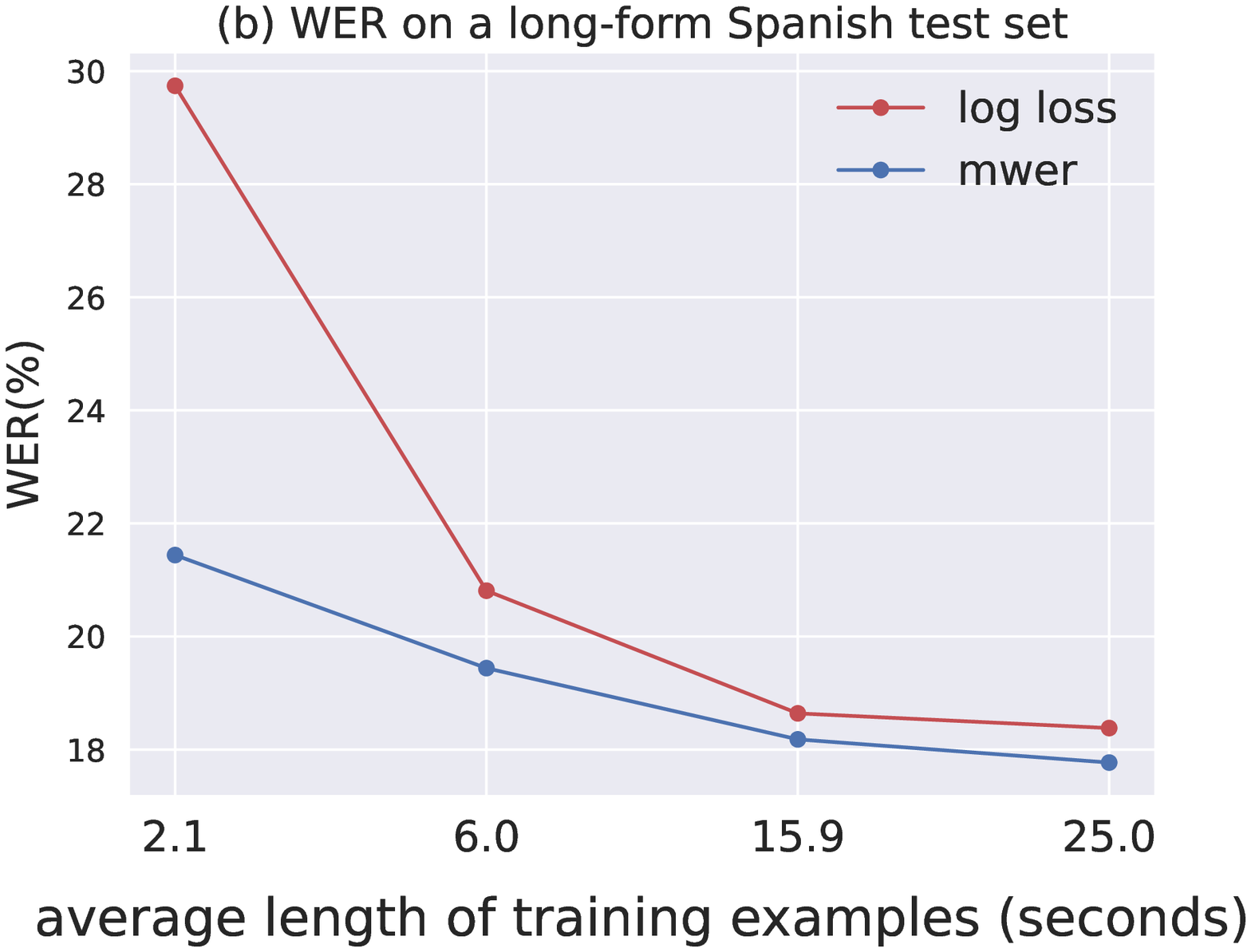} \\
    \end{tabular}
    \vspace{-0.1in}
    \caption{
    {\textbf{(a) E2E model fails to decode long audio}}. On SWBD concatenated sets, the WER  increases with the utterance length. 
    {\textbf{(b) We improve long-form speech recognition by increasing the length of training examples.}} On a long-form Spanish test set, the WER improves as we increase the training example length. The improvement is consistent for both log loss and \EMBR loss.} \label{fig:wer} 
\vspace{-0.2in}
\end{figure}


However, E2E models can fail miserably to decode long utterances~\cite{narayanan2019recognizing,chiu2021rnn}. We use a Conformer RNN-T model\footnote{The model is Conformer XL as in~\cite{zhang2021bigssl}. It performs well on Voice Search task.} to decode Switchboard rt03 set and 6 concatenated variants of the corpus. We group utterances of the same conversation and speaker ID, and sort them with increasing start timestamps. We concatenate $N$ utterances in the sorted order, where $N=2,4,8,16,32,$ all. We plot the word error rate (WER) and deletion errors versus the average utterance length on the 7 sets in Fig~\ref{fig:wer}(a). Note that they are derived from the same corpus, thus should have similar WERs. However, in Fig~\ref{fig:wer}(a) the WER, in particular deletion error, increases with the utterance length. It shows that E2E models are sensitive to input length, and can suffer from high deletions on long input.

E2E models are known to be sensitive to distribution mismatch and susceptible to overfitting.  In particular, if the model is trained on short utterances, it fails to generalize to long utterances~\cite{narayanan2019recognizing,chiu2021rnn}.
To improve E2E models robustness to long-form, ~\cite{chiu2021rnn} proposed various regularization techniques, and~\cite{narayanan2019recognizing} proposed to simulate long-form characteristics during training by manipulating models' LSTM states. In this work, we adopt a simpler but more direct solution: we train the model on longer segments while retaining the \emph{correct} acoustic context. We answer an important but overlooked question: is learning from short utterance optimal for E2E model?



We investigate the effect of training utterance length on the WER for long-form speech. The training data are long recordings of telephone calls.  
In addition to the text label, the transcription also contains annotated start and end times of the speech segments. This allows us to prepare the training set under different segmentation, by merging consecutive speech segments into a longer one to form training examples.
We retain non-speech audio between the speech segments. And a single training example can contain multiple speakers. See Fig.~\ref{fig:segment} for a demonstration. 


Furthermore, we compare two widely used training objectives for the RNN-T model on the long-form task, \ie the log loss (or RNN-T loss), and minimum word error rate (\EMBR) loss ~\cite{lu21b_interspeech,guo2020efficient,weng2019minimum,prabhavalkar2018minimum}. 
The log loss optimizes the log probability of the label sequence by marginalizing over alignments~\cite{graves2012sequence}. 
Despite an unbiased estimator in theory,
maximum likelihood training suffers from exposure bias~\cite{cui2021reducing,li21m_interspeech} in practice. The model makes predictions conditioned on the ground truth labels during training, but on the erroneous predicted labels during inference. This mismatch can hurt the model's generalization. 
On the other hand, the \EMBR loss directly minimizes the expected number of word errors, the evaluation metric at test time. \EMBR is a variant of edit-based minimum Bayes risk~\cite{shannon2017optimizing}, where the expectation is approximated with an empirical average of the $N$-best hypothesis from the beam search. Compared to the log loss, \EMBR loss is conditioned on not only the reference but also the competing hypotheses,
and its training procedure reduces the exposure bias by doing inference at training time. However, \EMBR training is more computationally expensive. 
\eat{Compared to the log loss, \EMBR uses sequence-discriminative criterion instead of cross-entropy,}

We conduct experiments on telephony datasets in four languages. Each language contains a few hundred hours of audio. Our main contributions are two-folds: Firstly, we show that increasing the length of training examples significantly improves the WER on the long-form task for both log loss and \EMBR loss. The average relative WER reduction is 15.7\% for log loss and 8.8\% for \EMBR loss. 
Fig.~\ref{fig:wer} (b) demonstrates the WER improvement on Spanish test set for both losses. \EMBR loss performs slightly better than the log loss, at the price of higher computational cost. The gain is more compelling when the training examples are short. 
Secondly, we propose a two-stage training recipe which achieves good WER with low computation costs: training first with log loss on \emph{long} utterances and then fine-tuning with \EMBR on \emph{short} utterances. 


\section{Method} \label{sec:method}
In this section, we describe the telephony datasets and how we apply segmentation to create longer training examples in \S\ref{subsec:segment}. We introduce the details of the log loss and \EMBR loss in \S\ref{subsec:loss}.

\subsection{Create long training examples}\label{subsec:segment}
\begin{figure}[tb]
    \centering
    \includegraphics[width=0.48\textwidth]{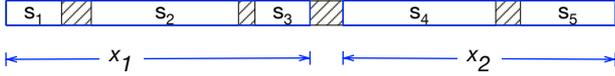}
    \caption{The training example is generated from segmentation of long audio. $s_i$ is transcribed speech segment, and the slash filled interval is non-speech audio. The training example $\vx_1$ contains speech segments $s_1, s_2$ and $s_3$ and the non-speech in-between. There can be speaker changes within one example.}\label{fig:segment}
\vspace{-0.2in}
\end{figure}

The datasets consist of recordings of telephone calls in 4 languages: Australian English (En), French (Fr), Mexican Spanish (Es), and Brazilian Portuguese (Pt). The total number of hours, and the average length of the audio recordings in seconds are summarized in Table~\ref{tab:data}. There are 2 telephony test sets in Es and Fr, and 1 test set in Pt and En. For both training and test sets, the audio is of a few minutes long. We also include an out-of-domain (OOD) test set collected from YouTube videos for each language, in order to monitor the generalization performance on non-telephony long-form speech.

\begin{table}[tb]
\vspace{-0.1in}
\centering
\caption{Datasets statistics: total hours of the dataset / the average length of the recordings in seconds.} \label{tab:data}
\vspace{-0.1in}
\begin{adjustbox}{max width=0.48\textwidth}
    \begin{tabular}{lccccc}
    \multirow{2}{*}{lang.}  & \multirow{2}{*}{train sets} & \multicolumn{3}{c}{test sets} \\
    \cmidrule{3-5}
    && tel 1 & tel 2 & OOD \\ \toprule
    Es &  552.1h / 130.4s & 25.3h / 101.8s & 24.7h / 184.5s & 9.7h / 512.3s \\
    Pt &  614.6h / 431.0s & 19.2h / 139.8s & - & 9.8h / 519.0s \\
    En & 215.1h / 62.2s & 19.3h / 130.4s& - & 6.4h / 493.2s \\
    Fr & 250.1h / 106.6s & 19.9h / 125.9s & 22.6h / 52.2s  &   10.0h / 611.9s \\ 
    \bottomrule
    \end{tabular}
\end{adjustbox}
\end{table}

\begin{table}[tb]
\thickmuskip=0mu
\medmuskip=0mu
\centering
\caption{Length of training examples (mean$\pm$std) in seconds under different segmentation regimes.} \label{tab:segment}
\vspace{-0.1in}
\begin{adjustbox}{max width=0.48\textwidth}
    \begin{tabular}{rcccc}
       seg.  & Es & Pt & En & Fr \\ \toprule
       raw & $2.1 \pm 2.9$ & $2.7 \pm 3.2$ & $3.5 \pm 4.7$ & $3.0 \pm 3.4$\\
       short & $6.0 \pm 4.1$ & $7.7 \pm 3.2$ & $7.3 \pm 5.3$ & $7.0 \pm 3.9$\\
       medium &$15.9 \pm 11.0$& $24.2 \pm 7.0$ & $14.4 \pm 10.8$ & $16.5 \pm 10.4$ \\
       long & $25.0 \pm 22.1$ & $49.4 \pm 14.7$ & $20.3 \pm 18.8$ & $24.7 \pm 20.0$\\\bottomrule
    \end{tabular}
\end{adjustbox}
\vspace{-0.2in}
\end{table}
Each minutes-long audio contains multiple annotated transcripts. Each transcript contains a tuple of (text label, start time, end time). In Fig.~\ref{fig:segment}, we refer to such a transcript tuple as a segment $s_i$. From the raw transcription segment $s_i$, we can sort them in order of starting time\footnote{The timestamp information is only used to segment the audio, but not used during training.} and group consecutive segments into a longer unit. For example, $s_1, s_2$ and $s_3$ are grouped together into one example $\vx_1$, and $s_4, s_5$ into $\vx_2$. We use these utterances of relatively longer length as training examples. We can control the length of the training examples by determining how we merge the raw segments. To retain the correct acoustic context, we keep the non-speech audio, like background noise and music, in-between speech segments in the training example, and there can be speaker change within one example. In principle, we can also use any voice activity detector (VAD)~\cite{thomas2015improvements,xu21b_interspeech} to segment the audio and control the length of training examples by varying the threshold of non-speech intervals. In our case, we simply use the segmentation provided in the annotation.

We prepare the training sets under 4 segmentation regimes, which we refer to as raw, short, medium, and long. Table~\ref{tab:segment} summarizes the mean and the standard deviation of the training example lengths under different regimes. For raw segmentation, we use the transcribed segments provided in the annotation without any merging. For short, medium, and long, raw segments are merged into longer utterances. From short to long, the number of segments that are merged, \ie the training example length, successively increases.

\subsection{Optimization objectives}\label{subsec:loss}
We denote the input utterance as $\vx$, and the token sequence as $\vy$. $\vy^*$ is the ground-truth target sequence. The RNN-T model outputs a probability over tokens for any pair of alignment. From the RNN-T output, we can derive the probability $\Pr(\vy|\vx)$ of any token sequence $\vy$ by marginalizing possible alignments using the forward backward algorithm~\cite{graves2012sequence}.

\newcolumntype{v}{>{\columncolor{LightRed}}c}
\newcolumntype{w}{>{\columncolor{LightBlue}}c}
\begin{table*}[tb]
    \centering
    \caption{WER (\%) under different training example lengths. The 3rd column is WERs of initialized models, the red block is for models trained with log loss, and blue block \EMBR loss. For both losses, the WER improves as the segmentation length increases. The average relative improvement is 15.7\% for log loss and 8.8\% for \EMBR.}
    \label{tab:wer_length}
    \vspace{-0.1in}
    \begin{tabular}{llcvvvvwwww}
 \rowcolor{white}   & & &\multicolumn{4}{c}{log loss} & \multicolumn{4}{c}{MWER loss} \\
\cmidrule(lr){4-7} \cmidrule(lr){8-11}
\rowcolor{white} 
\multirow{-2}{*}{lang.}  &  \multirow{-2}{*}{test set} & \multirow{-2}{*}{initialization} &
   raw & short & medium & long
   & raw & short & medium & long \\ 
\toprule
   \multirow{3}{*}{Es}
   & Telephony 1 &  34.0 
            & 29.7 & 20.8 & 18.6 & \bf{18.4} 
            & 21.4 & 19.4 & 18.2 & \bf{17.8} \\
   & Telephony 2 & 38.1
            & 33.9 & 25.9 & \bf{24.4} & 24.7 
            & 27.3 & 25.0 & 25.4 & \bf{24.1} \\ 
    & OOD YouTube & 17.6 
          & 18.4 & 15.9 & \bf{15.7} & 17.6  
            & 17.1 & 16.2 & 15.8 & \bf{15.4} \\  \midrule
  \multirow{2}{*}{Pt} 
  & Telephony  &  35.0 
        & 29.0 & 24.0 & 22.5 & \bf{21.8} 
         & 25.8 & 22.5 & 21.9 & \bf{21.7}\\
 & OOD YouTube & \bf{17.0}  
        & 17.5 & 21.9  & 18.0  & 17.5 
        & 17.1 &  18.0 & 18.0 & 17.1 \\ \midrule
 \multirow{2}{*}{En} 
 & Telephony  & 24.7 
        & 18.0  & 17.6 & 17.2 & \bf{17.0}
        & 17.3 &  17.2 & 17.0 & \bf{16.9} \\
 & OOD YouTube  & \bf{11.4} 
        & 12.3 & 12.0  & 11.8 & 11.5  
        & 12.0 & 11.8 & 11.7 & 11.6 \\
\midrule
 \multirow{3}{*}{Fr} 
 & Telephony 1  & 28.0 
            & 23.8 & 22.6 & 22.4 & \bf{22.4} 
            & 23.2 & 22.2 & 22.5 & \bf{22.1} \\
 & Telephony 2  &  29.3 
         & 24.5 & 24.4 & 24.0 & \bf{23.8} 
         & 24.0 & \bf{23.7} & 24.0 & 23.8 \\
 & OOD YouTube & 17.6 
    & 18.1 & 17.0 & \bf{16.7} & 18.6 
       & 18.0 &  17.2 & \bf{16.5} & 16.7\\ 
    \bottomrule
    \end{tabular}
    \vspace{-0.1in}
\end{table*}

\noindent\textbf{Log loss} is defined as the negative log probability of the ground-truth sequence. $\sL_{\text{ll}} = -\log \Pr(\vy^*|\vx).$


\noindent\textbf{\EMBR loss.}  We use $\ell(\vy, \vy^*)$ to denote the number of word errors in a hypothesis $\vy$ relative to $\vy^*$. Edit-based minimum Bayes risk (EMBR)~\cite{shannon2017optimizing} is defined as the expected number of word errors. We use the $N$-best hypotheses from beam search as empirical samples to approximate the expectation, following~\cite{weng2019minimum,prabhavalkar2018minimum,povey2005discriminative}. 
We denote the $N$-best list as $\text{Beam-}N(\vx)$. The \EMBR loss is defined as, 
$  \sL_{\text{mwer}} = \sum_{\vy} \Pr(\vy|\vx) \ell(\vy, \vy^*) 
 \approx \sum_{\vy_i \in \text{Beam-}N(\vx)} \widehat{P}(\vy_i) \left[ \ell(\vy_i, \vy^*) - \widehat{\ell} \right], 
$
where $\widehat{P}(\vy_i) = \frac{\Pr(\vy_i | \vx)}{ \sum_{\vy_i \in \text{Beam-}N(\vx)} \Pr(\vy_i | \vx)}$ is the re-normalized probability, and $\widehat{\ell} =  \sum_{\vy_i \in \text{Beam-}N(\vx)} \widehat{P}(\vy_i)\ell(\vy_i, \vy^*)$ is the average number of word errors of $N$-best hypotheses. \EMBR loss boosts the probability of the hypothesis that has better than average word errors, and reduces the probability of the one that is worse than the average. 
In practice, to stabilize \EMBR training~\cite{prabhavalkar2018minimum}, we interpolate the \EMBR loss with the log loss using a hyper-parameter $\lambda$, 
$\sL_{\lambda\text{-mwer}} = \sL_{\text{mwer}} + \lambda \sL_{\text{ll}}$.


There are two main differences between the log loss and \EMBR loss. 
The log loss only increases the probability of the ground-truth sequence, while the \EMBR performs discriminative training among competing hypotheses; The log loss takes the ground-truth history label as input to the prediction network of RNN-T, while the \EMBR takes the (potentially erroneous) prediction history label, which helps reduce exposure bias~\cite{cui2021reducing}.
\section{Experiment} \label{sec:exp}

\subsection{Setup}\label{subsec:setup}

\noindent\textbf{Model Architecture.}
We use the state-of-the-art RNN-T model~\cite{li2021better} with the Conformer encoder~\cite{gulati2020conformer}. The output tokens are 4,096 word-pieces. For the acoustic front-end, we use 128-dimensional log-Mel features, computed with a 32ms window and shifted every 10ms. The log-Mel features from 4 contiguous frames are stacked to form a 512-dimensional input, and then subsampled by a factor of 3.

\eat{To reduce the computational cost and memory footprint, we employed embedding decoder~\cite{botros2021tied} with the context equal to 2.} 

\noindent\textbf{Initialization model.} 
We first pre-train the RNN-T model on YouTube and multi-domain data, before training on the telephony data. For Fr, Es and Pt tasks, the initialization model is pre-trained on YouTube segments labeled by Rover ensemble teacher model predictions, and multi-domain data. Multi-domain data covers search, far-field, and telephony. Please refer to~\cite{doutre2021bridging} for more details. For En task, the initialization model is pre-trained on multi-domain data, the same as in~\cite{narayanan2019recognizing}. 

\eat{It covers domains of search, far-field, telephony and YouTube. The YouTube transcriptions are generated in a semi-supervised fashion~\cite{liao2013large}. }

All experiments are done on 8x8 Cloud TPU~\cite{jouppi2017datacenter} using the Tensorflow Lingvo toolkit~\cite{shen2019lingvo}.  For optimization, we use Adam optimizer~\cite{Kingma15} with batch size of 128. SpecAugment~\cite{park2019specaugment} is applied for both log loss and \EMBR training. For \EMBR, the number of hypothesis $N$ is 4, and $\lambda=0.03$ in $\sL_{\lambda\text{-mwer}}$.

\subsection{Improve long-form WER by training on long examples}\label{subsec:results}
\begin{figure}[tb]
    \centering
    \hspace{-0.1in}
    \includegraphics[width=0.35\textwidth]{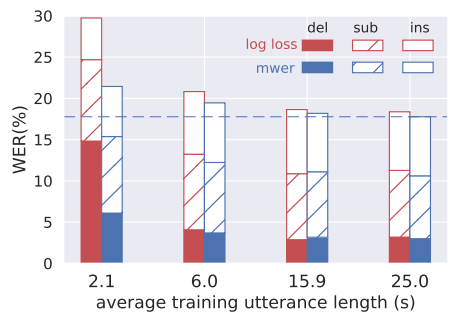}
    \caption{WERs on the Spanish Telephony 1 test set across different input lengths. The red bar is for log loss and blue for \EMBR. The blue dashed line marks the lowest WER of all models. When we train on the raw segment, the deletion error dominates the WER.  For both losses, the deletions reduce significantly as we increase the length of the training examples. \EMBR achieves the best WER, with slight gain in substitution error compared to the log loss.}\label{fig:breakdown}
    \vspace{-0.2in}
\end{figure}

In Table~\ref{tab:wer_length}, we report the WERs on test sets under different training example lengths. The length statistics of training example can be found in Table~\ref{tab:segment}, and the length statistics of test example is in Table~\ref{tab:data}. 
The 3rd column is the WER of the initialization checkpoint after pre-training. The red block is the WERs of the log loss models, and the blue block \EMBR loss. For both losses, the trend is that the longer the training example length, the better the WER. There is a 38\% relative WER reduction in Es Tel 1 set when trained with ``long'' segments compared to ``raw'' segments using log loss. The average relative gain across all in-domain test sets is 15.7\% for log loss and 8.8\% for \EMBR. ~\cite{doutre2021improving} observed a similar trend in a teacher-student training framework. If we compare WERs across the two losses, we find that WERs of \EMBR are lower than the log loss. When the training example is short, the benefit of \EMBR is more substantial, while the gain diminishes as the training example gets longer. ~\cite{li21m_interspeech} observed a similar trend for the oracle WER. We hypothesize that the lack of diversity in the $N$-best list for longer utterances~\cite{prabhavalkar2021less}, coupled with the use of a small $N$ impairs \EMBR training on long utterances.



To examine the source of mistakes of models predictions, we break down the word errors into deletions (del), substitutions (sub), and insertions (ins). We visualize the breakdown on Es Tel 1 test set in Fig.~\ref{fig:breakdown}. When we train on the raw segment with log loss, the del error dominates the WER, which takes up 50\% of the total errors. \EMBR model is better, but still suffers from relatively high deletions. A possible explanation is that the raw transcribed segment contains a minimum amount of non-speech audio, which makes the model less robust against noise and incurs high deletions. 
As we increase the length of the training examples, from an average of 2.1 seconds to 25.0 seconds, the deletions reduce significantly for both \EMBR and log loss, which as a result gives a much lower total WER. \EMBR achieves the best WER among all, with a slight gain in substitutions compared to the log loss. 


On the out-of-domain long-form YouTube set, increasing input length can improve WER in some cases, like Spanish and French, with the improvement from reduced deletions. But the gain is not always consistent, because of overfitting to the telephony domain. 

Lastly we evaluate the WER on short-form data. Moreover, we want to verify whether doing inference on long-form speech improves over the segmentation-then-inference approach. We segment the En OOD YouTube set into short utterances. The segmentation is provided by human annotation, and the average length is 11.1 seconds. We evaluate WER on the segmented test set, and compare it with the long-form decoding result (Table~\ref{tab:wer_length} row 9) in Table~\ref{tab:shortform}. Decoding on long audio has lower WERs, with improvement in all del/sub/ins. Besides, short-form WER is relatively stable across different training example lengths. This reassures us that training with longer utterances does not hurt the short-form performance. 

\begin{table}[tb]
\centering
    \caption{WER (\%) of decoding on long-form \vs on the segmented short utterances in En OOD dataset. Long-form inference is better.}\label{tab:shortform}
    \vspace{-0.1in}
    \begin{tabular}{rccccc}
\multirow{2}{*}{loss} & \multirow{2}{*}{\thead{testing \\ segment}} & \multicolumn{4}{c}{training segment}   \\
\cmidrule{3-6}
& & raw & short & medium & long \\ \toprule
  \multirow{2}{*}{log loss} & long  & 12.3 & 12.0  & 11.8 & 11.5  \\ 
                        & short  & 14.5 & 14.0 & 14.4 & 13.9 \\ \midrule
  \multirow{2}{*}{MWER loss} & long & 12.0 & 11.8 & 11.7 & 11.6 \\ 
                        & short &  14.1 & 14.0 & 13.9 & 14.2   \\ \bottomrule
    \end{tabular} 
\end{table}

\subsection{Best of both worlds: efficient two-stage training recipe}\label{subsec:twostage}



Despite good WERs, \EMBR training is computationally expensive, and it has diminished gains over the log loss when the input length increases. To this end, we experiment with a two-stage training recipe. We first train with log loss on \emph{long} segments, and then fine-tune with \EMBR loss on \emph{raw} segments. We hope to enjoy the best of both worlds: have the fast training speed and the benefit from longer inputs by log-loss, and have the good WER of the \EMBR loss. 
For the \EMBR fine-tuning, we experiment with both fine-tuning the full model, and fine-tuning the decoder only, \ie the prediction network and the joint network of RNN-T. The intuition behind it is that we train the model, especially the encoder, to capture the long-form characteristics by log loss in the 1st stage. And in the 2nd stage, we fine-tune the decoder to output better predictions with \EMBR loss. In the two-stage experiment, we set the $\lambda$ in $\sL_{\lambda\text{-mwer}}$ to be smaller. $\lambda=0.003$ for full model fine-tuning, and 0 for decoder fine-tuning.

\begin{table}[tb]
    \centering
    \caption{WER (\%) of the two-stage training. The 2nd stage \EMBR fine-tuning always improves over the 1st stage, and gets close or even outperforms (in bold) the best WER from Table~\ref{tab:wer_length}.}
    \label{tab:2stage}
    \vspace{-0.1in}
    \begin{tabular}{llggcc}
\rowcolor{white}\multirow{2}{*}{lang.}  & \multirow{2}{*}{test} &   &   & \multicolumn{2}{c}{2nd stage (MWER)}    \\
\cmidrule{5-6}
\rowcolor{white} & & \multirow{-2.5}{*}{\thead{best of\\Table~\ref{tab:wer_length}}} & \multirow{-2.5}{*}{\thead{1st stage \\ (log loss)}} & full mdl & dec only \\
\toprule
 \multirow{3}{*}{Es} 
 & Tel 1 &  17.8 & 18.4 & 17.9 & \bf{17.5} \\
 & Tel 2 &  24.1 & 24.7 &  24.4 & 24.4 \\
 & OOD &  15.4 & 17.6 & 18.4 & 17.0 \\ 
 \midrule
 \multirow{2}{*}{Pt}
 & Tel 1 & 21.7 & 21.8 & \bf{21.6} & 21.8 \\
 & OOD & 17.0 &17.5 & 17.4 & 17.4 \\ 
\midrule
  \multirow{2}{*}{En}
 & Tel 1 & 16.9 & 17.0 & 16.9 & 16.9 \\
 & OOD & 11.4 & 11.5 & 11.5 & 11.4 \\ 
  \midrule
  \multirow{3}{*}{Fr} 
 & Tel 1 &22.1 & 22.4 & \bf{21.9} & 22.0 \\
 & Tel 2 & 23.7 & 23.8 & \bf{23.3}  & 23.5 \\
 & OOD & 16.5 & 18.6 & 17.0 & 17.1 \\ \bottomrule
    \end{tabular}
\end{table} 

\begin{table}[tb]
    \centering
    \caption{Computation costs of different training recipes on the Spanish task. The total cost of two-stage training with decoder fine-tuning is 13.4 hours, $1/4$ of the cost of \EMBR training on long segments.}
    \label{tab:cost}
    \vspace{-0.1in}
    \begin{tabular}{rcccc}
   & \multirow{2}{*}{\EMBR}  &  \multirow{2}{*}{\thead{log loss \\ (1st stage)}} & \multicolumn{2}{c}{2nd stage (MWER)}    \\ 
   \cmidrule{4-5}
   & &  & full mdl & dec only \\ \toprule
   segmentation & long & long & raw & raw \\ \midrule
    seconds / step & 3.88 & 2.67 &  1.35 & 1.08 \\
    \# steps  & 60k & 8k & 50k & 16k \\
    total hours & 44.5 & 8.6 & 18.7 & 4.8 \\ \bottomrule
    \end{tabular}
    \vspace{-0.1in}
\end{table} 
\eat{achieved by one stage training }

In Table~\ref{tab:2stage}, the numbers in gray block are copied from Table~\ref{tab:wer_length}: the 3rd column is the best WER on each test set;
the 4th column is the WER of 1st stage log loss training, column 7 of Table~\ref{tab:wer_length}. The 5th and 6th columns are the WERs after \EMBR fine-tuning on raw segments, w.r.t. all weights and decoder weights respectively. On in-domain telephony test sets, fine-tuning with \EMBR consistently improves over the 1st stage WER. It gets close or even outperforms the best WER in some test sets. 

Table~\ref{tab:cost} compares the computation costs of different training recipes. We break down the training time into two parts: average seconds per training step, and the number of steps until the best test WER. The last row is the total training time in hours\eat{, which is the product of the previous two rows}. 
Comparing \EMBR with log loss on long segments in the 2nd and 3rd columns, \EMBR takes longer to train. It is slower every step, as it performs beam search and error computation; It also takes more steps to converge. The 2nd stage \EMBR fine-tunes on raw segments, which is 3 times faster than on long segments. Since the number of parameters in the decoder is only $1/10$ of the encoder, fine-tuning on decoder converges in much fewer steps than the full model. Thus it reduces the \EMBR fine-tuning time from 18.7 hours to 4.8 hours. To summarize, first training with log loss on \emph{long} segments and then fine-tuning with \EMBR on \emph{raw} segments can achieve good WERs with relatively low computation costs.

\eat{It gets even worse when the training example is longer.}



\section{Conclusion}
\label{sec:conclude}

In this work, we improve RNN-T and \EMBR training on long-form telephony speech recognition by increasing lengths of training utterances. The average relative WER reduction is 15.7\% for log loss and 8.8\% for \EMBR loss. We propose an efficient two-stage training recipe which achieves good WER with low computation costs. Future works include improving \EMBR training on long utterances, and incorporation of an external language model to improve WER.



 \section{Acknowledgement}
We are grateful to Chung-Cheng Chiu, Wei Han, Yu Zhang, Pedro Moreno Mengibar, Arun Narayanan, Qiujia Li, Yongqiang Wang, Ruoming Pang, Hank Liao, Basi Garc\'{i}a, Han Lu, Qian Zhang, Hasim Sak, Oren Litvin for their help and suggestions.

\section{REFERENCES}
\label{sec:refs}
\renewcommand{\section}[2]{}
\setlength{\bibsep}{2pt plus 0.6ex}
\bibliographystyle{IEEEbib}
\bibliography{refs}

\end{document}